\documentclass[
reprint,
amsmath,
amssymb,
aps,
floatfix
]{revtex4-2}

\usepackage{graphicx}
\graphicspath{ {./figures/} }
\usepackage{dcolumn} 
\usepackage{bm} 
\usepackage{array} 
\newcolumntype{P}[1]{>{\centering\arraybackslash}p{#1}} 
\usepackage{braket}
\usepackage{dsfont} 

\usepackage{xcolor}
\usepackage{hyperref}
\hypersetup{
breaklinks=true,
colorlinks=true,
citecolor=black,
urlcolor=black,
linkcolor=black,
}

\begin{document}

\title{Cycle discrete-time quantum walks on a noisy quantum computer}

\author{Vivek Wadhia\textsuperscript{1}}
\author{Nicholas Chancellor\textsuperscript{1}}
 \email{nicholas.chancellor@durham.ac.uk}
\author{Viv Kendon\textsuperscript{1,2}}
 \email{viv.kendon@strath.ac.uk}

\affiliation{\mbox{\textsuperscript{1}Department of Physics, Durham University, Durham DH1 3LE, United Kingdom} \\
\mbox{\textsuperscript{2}Department of Physics, University of Strathclyde, Glasgow G4 0NG, United Kingdom} \\}

\date{\today}

\begin{abstract}
The rapid development of quantum computing has led to increasing interest in quantum algorithms for a variety of different applications. Quantum walks have also experienced a surge in interest due to their potential use in quantum algorithms. Using the qiskit software package, we test how accurately the current generation of quantum computers provided by IBM can simulate a cycle discrete-time quantum walk. Implementing an 8-node, 8-step walk and a simpler 4-node, 4-step discrete-time quantum walk on an IBM quantum device known as \texttt{ibmq\_quito}, the results for each step of the respective walks are presented. A custom noise model is developed in order to estimate that noise levels in the \texttt{ibmq\_santiago} quantum device would need to be reduced by at least 94\% in order to execute a 16-node, 16-step cycle discrete-time quantum walk to a reasonable level of fidelity.   
\end{abstract}

\maketitle

\section{\label{sec:level1}Introduction}

In recent years, quantum computing \cite{Feyn1,Deutsch} has come to the forefront of developments in physics,
with the promise of eventually being able to perform certain computations more efficiently than on a classical computer \cite{Josza1,Bernstein1,Simon1}. This has motivated the creation of a number of well-known quantum algorithms over the years in order to utilise this computational speed-up, such as Shor's algorithm \cite{Shor_1997} and Grover's algorithm \cite{Grover1}.

Along with the development of quantum computing, quantum walks \cite{Gudder1, Aharonov_1} have also gained increasing interest. In large part, this is due to their application to quantum algorithms \cite{Farhi_1, DAharonov1, Ambainis1, Shenvi1}. Quantum walks come in two main types, continuous-time and discrete-time.  In this work, we focus on discrete-time quantum walks
(DTQW), in particular a DTQW on a cycle graph \cite{Venegas1}, because it has a convenient encoding when executed on a digital quantum computer.
However, current hardware has significant imperfections.  Our aim in this work is to determine how much IBM processors need to improve to run DTQWs on cycle graphs.

This paper is structured as follows. In section \ref{sec:background}, an implementation of a cycle discrete-time quantum walk is presented using a binary encoding. In section \ref{sec:execution}, we present the results of executing this algorithm for eight steps of an 8-node quantum walk, and then for four steps of a 4-node quantum walk, on an IBM quantum device known as \texttt{ibmq\_quito}, as well as on a local, noiseless simulation using a quantum device known as \texttt{qasm\_simulator}. 
The results for both devices are compared using a fidelity measure for probability distributions. In section \ref{sec:custom_noise_models}, the same algorithm is extended to 
16 steps of a 16-node DTQW and run on \texttt{ibmq\_santiago}, whilst also being run on a noisy version of the simulator, for varying noise levels, with the fidelity being calculated for every step of the walk. Results comparing the fidelity of the \texttt{ibmq\_santiago} device with simulated noisy walks are presented. We conclude in section \ref{sec:conclusion}. 

\section{Background}
\label{sec:background}

\subsection{Quantum gates}
\label{sec:quantum_gates}

In analogy with classical logic gates, quantum computations can be constructed from quantum logic gates. In this work we use the following quantum gates: $X$ gate acts in the same way a classical logical NOT gate does, $\ket{0} \rightarrow \ket{1}$ and $\ket{1} \rightarrow \ket{0}$. The Hadamard ($H$) gate creates a superposition state, $\ket{0} \rightarrow \frac{1}{\sqrt{2}}\left( \ket{0} + \ket{1} \right)$ and $\ket{1} \rightarrow \frac{1}{\sqrt{2}}\left( \ket{0} - \ket{1} \right)$. A \textit{CNOT} (Controlled-Not) gate flips the state of the target qubit if the state of the control qubit is $\ket{1}$. Similarly, a Toffoli (\textit{C3-X}) gate flips the state of the target qubit if both of the control qubits are in the state $\ket{1}$. \textit{C4-NOT} (\textit{C4-X}) and \textit{C5-NOT} (\textit{C5-X}) gates flip the state of the target qubit if all 3 or 4 of the respective control qubits are in the state $\ket{1}$. An $SX$ ($\sqrt{X}$) gate acts $\ket{0} \rightarrow \frac{1}{\sqrt{2}} \left(\ket{0} -i\ket{1} \right)$ and $\ket{1} \rightarrow \frac{1}{\sqrt{2}} \left(\ket{0} +i\ket{1} \right)$, where $i$ is the imaginary unit (${i}^{2} = -1$). The identity gate (\textit{ID}) leaves any state it acts on unchanged. $RZ$ performs a rotation of the qubit state around the z-axis by a given angle.

For all DTQW quantum circuits used, the $H$ gate acts as the Hadamard coin operator. $X$, \textit{C-NOT}, Toffoli, \textit{C4-NOT} and \textit{C5-NOT} gates all act to change the value of the binary string (quantum register) that denotes the position of the walker.

\subsection{Quantum walk on a graph}
\label{sec:QW_on_a_graph}

A graph $g(v,e)$ is defined as a set of nodes $v$ and edges $e$ that connect pairs of nodes. Of particular interest for this work are cycle graphs, denoted by $C_{n}$ where $n$ is the number of nodes. For example, figure \ref{fig:8_node_graph} is a $C_{8}$ graph with the nodes labelled as used in the quantum walk algorithms.

A discrete-time quantum walk on a cycle graph requires a quantum walker that moves on the nodes of a graph, accompanied by a coin that is ``flipped'' to determine the direction to move in \cite{Venegas1,Kendon1}. The full quantum system (of the walk) is a combination of the position state and the coin state, the basis states are written $\ket{x,c}$ where $x$ labels the node and $c\in\{0,1\}$ the coin state. The general state of a DTQW at a time $t$ is
\begin{equation}
    \ket{\psi(t)} = \sum_{x,c} \alpha_{x,c}(t) \ket{x,c},
\end{equation}
where $\alpha_{x,c}(t) \in \mathbb{C}$ and normalized as $\sum_{x,c}|\alpha_{x,c}(t)|^2 = 1$.
Conventionally, the walker is initialised at the origin, i.e. $x=0$ at $t=0$, and the coin is in some chosen initial state.
The coin is ``flipped'' by applying a unitary operator, known as a coin operator.  For this work we use
\begin{equation}
    \hat{\bf{C}} = \frac{1}{\sqrt{2}} \begin{pmatrix}
                                 1 && 1 \\
                                 1 && -1
                                 \end{pmatrix},
\end{equation}
which is also known as a Hadamard coin, since the matrix is the same as a Hadamard quantum gate. A conditional shift operator then acts to move the quantum walker to adjacent nodes of the graph. This requires a mapping between coin states and the nodes at the ends of the edges to be specified \cite{Kendon2}
\begin{equation}
\label{eqn:mapping}
    \zeta : \mathbb{Z}_{|v|} \times \mathbb{Z}_{d} \rightarrow \mathbb{Z}_{|v|} \times \mathbb{Z}_{d} : (x,c) \mapsto \zeta(x,c) = (y,k).
\end{equation}
$\mathbb{Z}_{n}$ is the additive group of integers $\{0,\ldots,n-1\}$ modulo $n$, $d$ is the degree of the node (the number of edges connecting to it), $x$ and $y$ label each end of edge $(x,y)$ and $k$ is the updated coin state. The shift then acts
\begin{equation}
    \hat{\bf{S}} \ket{x,c} = \ket{y,k}.
\end{equation}
In other words, moving the walker and coin along the edge ($x,y$), with $x$ being the starting node, $y$ being the end node and updating the coin state, according to the mapping in equation \eqref{eqn:mapping}. The application of the coin operator followed by the shift operator, at each unit time step, is sometimes denoted by the unitary operator $\hat{\bf{T}}$, i.e.,
\begin{equation}
    \hat{\bf{T}} = \hat{\bf{S}} \hat{\bf{C}} .
\end{equation}
Therefore, a DTQW of $t$ time steps is achieved by applying $\hat{\bf{T}}$, $t$ times,
\begin{equation}
    \ket{\psi(t)} = \hat{\bf{T}}^{t} \ket{\psi(0)}.
\end{equation}
For this work the quantum register is always initialised in the all zero state. That is $\ket{0000}$, $\ket{000}$ and $\ket{00000}$ for the 8-node, 4-node and 16-node DTQWs respectively.

\subsection{Algorithm}
\label{sec:8_node_algorithm}

\begin{figure}[!htb]
\includegraphics[scale=0.3]{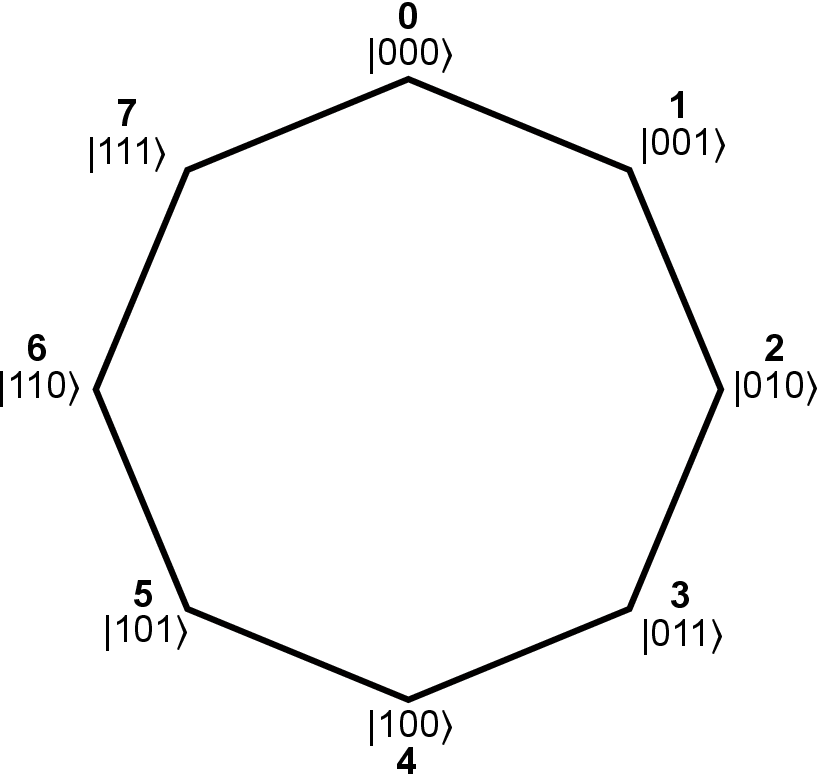}
\centering
\caption{8-node cycle graph. The binary string (i.e.~the state of the quantum register) gives the site number shown in decimal (bold).}
\label{fig:8_node_graph}
\end{figure}

Using the work of \citeauthor{GithubQW} \cite{GithubQW}, who in turn build on the work of \citeauthor{douglas1} \cite{douglas1}, it is possible to build a quantum circuit that performs a quantum walk on a cycle graph. For our purposes, we consider the example presented in \cite{GithubQW}, namely, that of a 4-qubit circuit that corresponds to an 8-node cycle graph. The cycle graph is depicted in  figure \ref{fig:8_node_graph}. The nodes of the graph correspond to the possible positions of the walker.

\begin{figure}[!htb]
\centering
\includegraphics[width = \columnwidth]{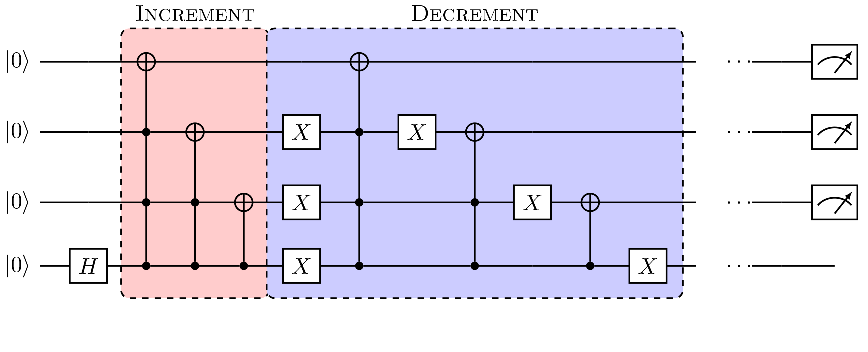}
\caption{ Quantum circuit of \cite{GithubQW}. Figure depicts one step of the walk. $H$, INCREMENT and DECREMENT gate sets are repeated for the desired number of steps. Quantum gates as they appear (from left to right) are: Hadamard, \textit{C4-X}, Toffoli, \textit{CNOT}, \textit{X} etc., ending with measurement.  Bottom most qubit (as pictured in the diagram) acts as the coin, the rest of the qubits are used to determine the position label (see figure \ref{fig:8_node_graph}).}
\label{fig:8node_circuit}
\end{figure}

The circuit for this 8-node walk is shown in figure \ref{fig:8node_circuit}. The quantum register consists of four qubits, three of which are used to represent the position of the walker as a binary label, while the final qubit is used as the coin state. Essentially, the circuit is built of two large sets of gates, an INCREMENT step gate and a DECREMENT step gate, acting as the shift operator, which themselves consist of \textit{CNOT} and $X$ gates. These INCREMENT and DECREMENT gates are repeated depending on the number of steps required. The coin operator used is a Hadamard gate applied before each INCREMENT step.  One step involves $H$, INCREMENT, DECREMENT; two steps involve $H$, INCREMENT, DECREMENT, $H$, INCREMENT, DECREMENT, and so forth. 

\newpage

\subsection{Fidelity}\label{ssec:fidelity}

In order to make a quantitative comparison between results from two different processors, such as \texttt{ibmq\_quito} and the \texttt{qasm\_simulator}, we now introduce the measure of fidelity to be used. The Hellinger fidelity \cite{fidelity} is defined as
\begin{equation}\label{eq:fidel}
    \mathcal{F}_{\mathcal{H}} = {( 1-\mathcal{{H}}^{2} )}^{2}
\end{equation}
where $\mathcal{H}$ is the Hellinger distance.
\begin{equation}
    \mathcal{H} = \left[ \int \left( \sqrt{p(z)} - \sqrt{q(z)} \right)^{2} dz \right]^\frac{1}{2}
\end{equation}
where $p(z)$ and $q(z)$ are two probability distributions \cite{Hellinger_distance_lecture_notes}.
Essentially, the Hellinger fidelity is a means of comparing how similar two classical probability distributions are. The probability distributions we compare are the results of the executed circuit which are expressed as counts, calculated from the number of times each position is measured as the outcome in a set of repeated runs of the circuit. The fidelity takes values in the range 0 to 1, with a value of 1 indicating a perfect correlation between distributions, and a value of 0 indicating a perfect anti-correlation between distributions. A value of 0.5 indicates no correlation, the distributions have a random overlap. This means we expect a fidelity of around 0.5 when we compare completely noisy states to the perfect outcome of a noise-free simulation. Noisy states tend to a value of 0.5 because this indicates maximum randomness in the results. A fidelity of less than 0.5 indicates the results of the computation are incorrect and, on average, anti-correlated. A fidelity of 0.8 or higher shows a clear correlation between the distributions and therefore indicates the results of the computation are more often correct than incorrect over repeated runs.

\subsection{Backends and transpilation}
\label{sec:transpile}
A \textit{backend} refers to either a simulation of a quantum device or a real quantum device provided by IBM. There are three backends used in this work. The first is \texttt{qasm\_simulator}, a local, noiseless, error-less, classical simulation of an arbitrary 5-qubit quantum device. The second is \texttt{ibmq\_quito}, a real 5-qubit quantum device, and the third is \texttt{ibmq\_santiago}, also a real 5-qubit quantum device. Quantum circuits are executed by first being programmed using the qiskit software package, and then being submitted online (for the case of real quantum devices) to join a queue of all jobs to be executed on that particular backend. Once execution is completed, results are displayed through the qiskit software frontend. 

\textit{Transpilation} \cite{transpile} is the process of converting quantum circuits, consisting of a range of standard gates into the native gates used by IBM quantum devices. There are only certain gates that can be executed on IBM quantum hardware. The gates are: \{\textit{CNOT}$, ID, RZ, SX, X$\}. These form a universal set of gates, so all operations can be implemented using only these gates. The transpilation process occurs automatically when the user submits the circuit to be executed to the backend. Transpilation does not occur in the \texttt{qasm\_simulator} as it is not a real quantum device, and can simulate all gates directly. 

\subsection{Software versions used}

Design and execution of circuits was done using \verb|qiskit_v0.24.0|~running on \verb|Python 3.8.5|. \verb|Spyder 4.1.5| is used as the IDE to write code along with \verb|IPython 7.19.0|. Table \ref{table:qiskit_versions} shows the version for each qiskit module.

\begin{table}[!htb]
\begin{tabular}{ | P{4cm} | P{1.5cm}| P{1cm} | } 
  \hline
  \underline{Module} & \underline{Version} \\ 
  \hline
  \verb|qiskit-terra| & \verb|0.16.4| \\ 
  \hline
  \verb|qiskit-aer| & \verb|0.7.6| \\ 
  \hline
  \verb|qiskit-ignis| & \verb|0.5.2| \\ 
  \hline
  \verb|qiskit-ibmq-provider| & \verb|0.12.1| \\
  \hline
  \verb|qiskit-aqua| & \verb|0.8.2| \\
  \hline
\end{tabular}
\caption{\centering Qiskit module versions.}
\label{table:qiskit_versions}
\end{table}

\newpage

\section{Runs and Simulations}
\label{sec:execution}
 
The rest of this paper examines DTQWs in the following order of dimensionality. Beginning with 8-node, followed by 4-node, and finally in section \ref{sec:custom_noise_models} a 16-node. The reason for this ordering is to determine the impact of noise on the fidelity of results from both smaller and larger systems, as compared to the 8-node system already established in literature.  

\subsection{Eight node quantum walk}
\label{sec:8_node_walk}
\begin{figure}[!htb]
\centering
\includegraphics[width=\columnwidth]{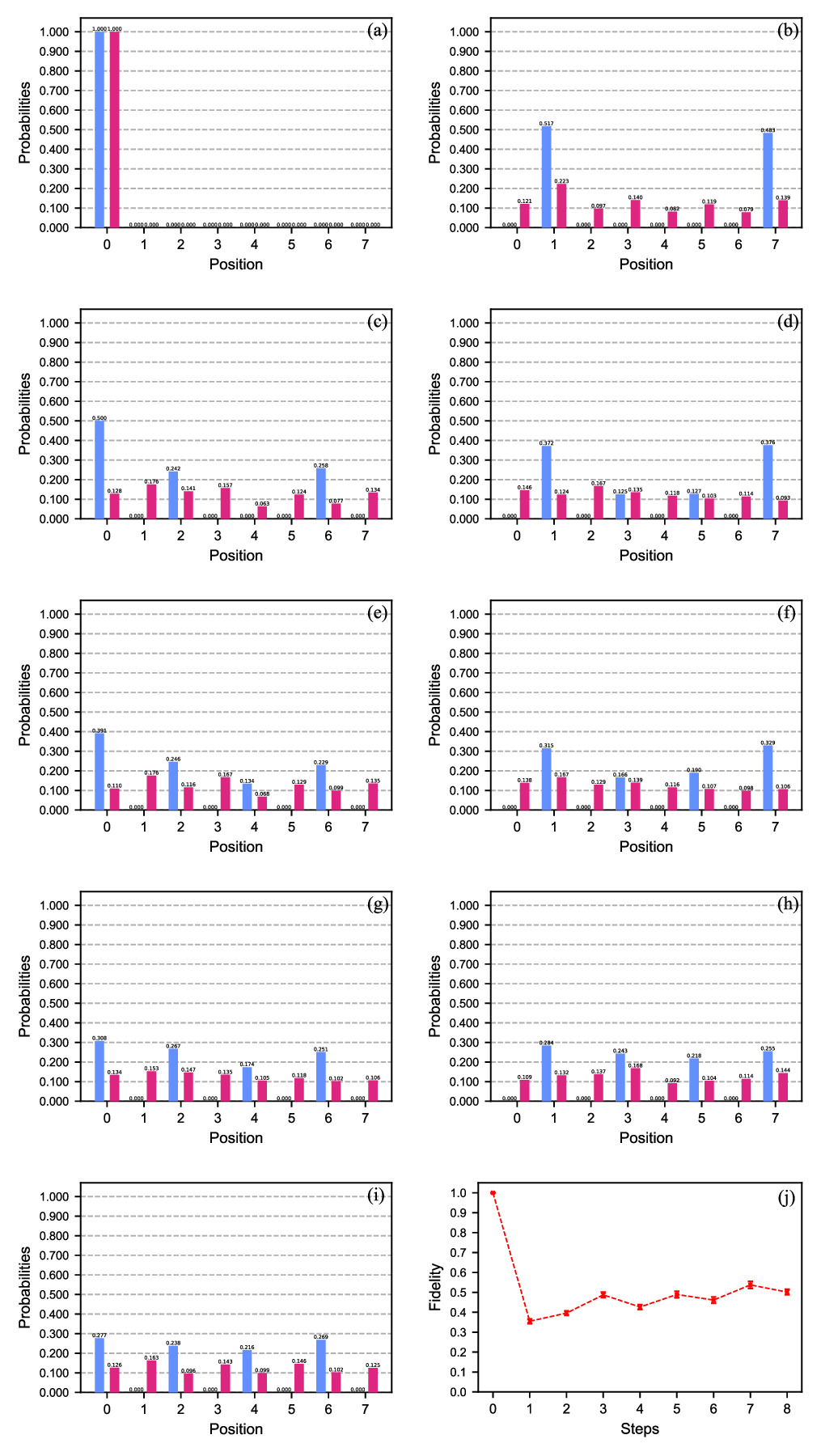}
\caption{Discrete-time quantum walk for an 8-node graph. Panels (a) to (i) show steps 0 to 8 of the walk respectively. Bars represent the probability of finding the walker at a particular site. The blue bars represent the result of the simulator i.e. the expected result and red bars the real device. Panel (j) shows the  Hellinger fidelity \eqref{eq:fidel} between the real device and the simulator for each step of the walk (dotted line is to guide the eye). Error bars on panel (j) represent standard error.}
\label{fig:8node_DTQW_result}
\end{figure}

Using the algorithm and circuit presented in section \ref{sec:8_node_algorithm} and figure \ref{fig:8node_circuit} respectively, an eight step DTQW on a cycle graph consisting of eight nodes is executed on \texttt{ibmq\_quito} and the \texttt{qasm\_simulator}, with the counts for each measured location of the quantum walk recorded for each step of the walk. To do this, the walk needs to be repeated for each number of steps, i.e., the walk is executed for a specific number of steps and the register is measured. Then the walk is restarted, the desired number of steps are executed, and the register is again measured. This process is repeated to obtain results for a all the diferent steps in a walk. To compare the performance of the \texttt{ibmq\_quito} versus the \texttt{qasm\_simulator}, the fidelity defined in section \ref{ssec:fidelity} is calculated from the probability distributions derived from the repeated runs.  Since the simulator is regarded as providing the ideal (expected) result, free of the effect of any noise or errors, the fidelity quantifies the performance of the real \texttt{ibmq\_quito} hardware.
The results are presented in figure \ref{fig:8node_DTQW_result}.

As can be seen in figure \ref{fig:8node_DTQW_result}, the \texttt{ibmq\_quito} device performs poorly. A basic expectation would be that the real device follows the first few steps of the walk with a reasonable level of fidelity. However, panel (j) of figure \ref{fig:8node_DTQW_result} shows that the performance of the device does not achieve this, with an average fidelity of 0.45 from step one onwards. Furthermore, a fidelity below 0.5 indicates an anti-correlation between the expected results and the real device. This could be due to bit flip error(s) occurring early in the computation, causing the value of the quantum register to become anti-correlated compared to the expected value. However, as the number of steps increase, the fidelity tends to 0.5, indicating the dominance of noise for longer computations.
\newpage
\clearpage

\subsection{Four Node Quantum Walk}
\label{sec:4_node_walk}

\begin{figure}[!htb]
\includegraphics[scale=0.3]{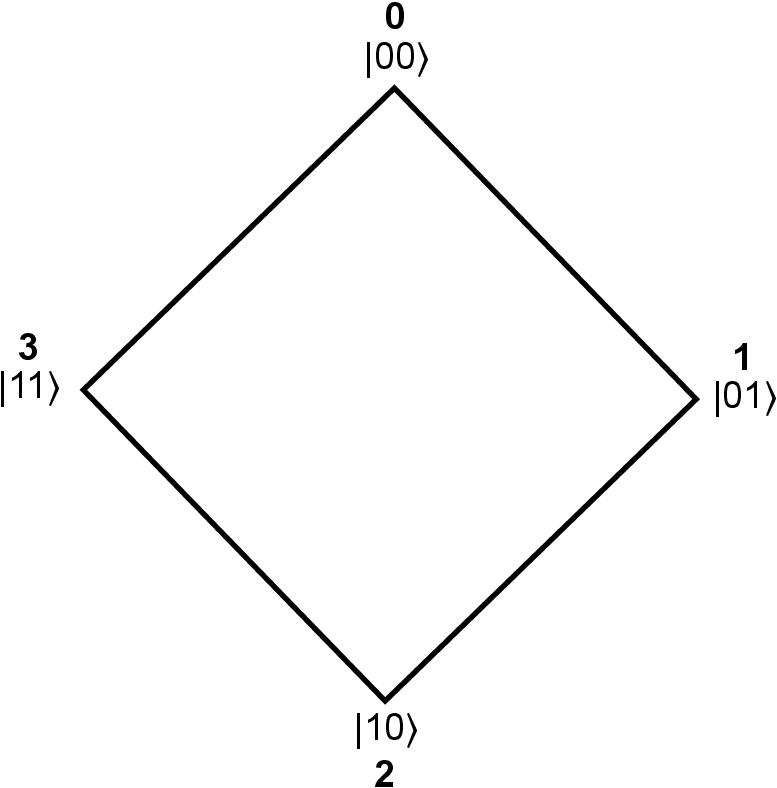}
\centering
\caption{4-node quantum walk graph. As in figure \ref{fig:8_node_graph} nodes represent the possible positions of the walker and the binary string corresponds to the label number in decimal (bold).}
\label{fig:4node_DTQW_graph}
\end{figure}

Let us now consider a simpler version of the algorithm presented in section \ref{sec:8_node_algorithm}, one that consists of only four nodes instead of eight, as shown in  figure \ref{fig:4node_DTQW_graph}.

\begin{figure}[!htb]
\centering
\includegraphics[width = 7.6cm, height = 2.15cm]{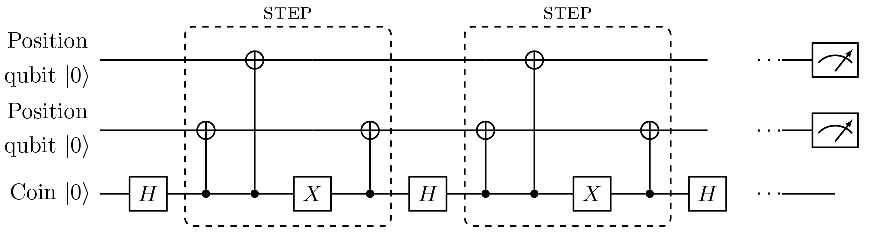}
\caption{Circuit for 4-node quantum walk. Decomposition of STEP gates are enclosed by the dashed lines. Coin operation is given by a single Hadamard gate acting on qubit three. Two complete steps of the quantum walk are shown, these gates are repeated for the desired number of steps.}
\label{fig:4node_DTQW_circuit}
\end{figure}

The circuit now needs just three qubits in total, qubits one and two are used to record the position and qubit three is used for the coin. This quantum walk simulation circuit is shown in figure \ref{fig:4node_DTQW_circuit}. Specifically, the pattern the circuit follows is: $H$ acting on coin, STEP, $H$ acting on coin, STEP and so on for further steps of the quantum walk. The circuit only consists of one large gate, a STEP gate that incorporates the functionality of the INCREMENT and DECREMENT gates of \cite{GithubQW}.

\begin{figure}[!htb]
\includegraphics[width=\columnwidth]
{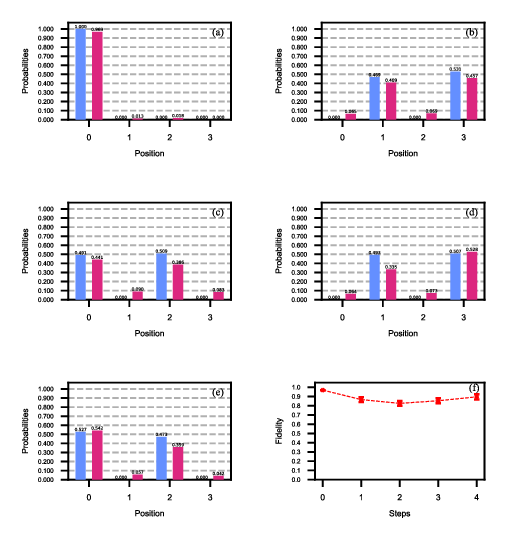}
\caption{DTQW for a 4-node graph. Panels (a) to (e) show steps 0 to 4 of the walk respectively. Bars represent the probability of finding the walker at a certain site denoted by a bit string. The blue bars represent the result of the simulator and red bars the result of the real device. Panel (f) shows the fidelity, comparing the two probability distributions, of the walk for each step between the real device and simulator. Error bars represent standard error.}
\label{fig:4_node_DTQW_panels}
\end{figure}

This algorithm was executed on \texttt{ibmq\_quito} for four steps of the walk, with the counts measured at each step and the fidelity with the \texttt{qasm\_simulator} calculated. The results are presented in figure \ref{fig:4_node_DTQW_panels}.

As can be seen in figure \ref{fig:4_node_DTQW_panels}, \texttt{ibm\_quito} performs the 4-node quantum walk with far higher fidelity than the 8-node. Although the fidelity of the results do decrease slightly during the initial steps, all steps are maintained at a fidelity of higher than 0.8, showing a good correlation between simulator and the real device. In other words, the correct computations are being performed, on average. This is in clear contrast to the results presented in figure \ref{fig:8node_DTQW_result}.

\subsection{Analysis of 8- and 4- node walks }
\label{sec:8_4_walk_analysis}

Let us now consider these results, and why the 4-node quantum walk performs similarly to the simulator whilst the 8-node counterpart does not. On inspection of their respective quantum circuits (shown in  figure \ref{fig:4node_DTQW_circuit} and figure \ref{fig:8node_circuit} respectively), both appear to be relatively simple. However, focusing on the INCREMENT and DECREMENT gates of the 8-node circuit, they contain Toffoli and \textit{C4-NOT} (\textit{C4-X}) gates. Implementation of these gates proves to be a non-trivial task. The qiskit software has a method of implementing Toffoli gates by reducing them into a set of other gates. In total, this decomposition consists of 18 gates: six \textit{CNOT} gates, 10 $RZ$ gates and 2 $SX$ gates for each Toffoli. This is already a large increase in the total number of gates required to perform a single Toffoli operation. For the case of a \textit{C4-NOT} or generally any C$n$-NOT gate of higher order (e.g.~\textit{C5-X}, \textit{C6-X} etc$.$), there currently exists no optimised way of implementing these gates in qiskit. A \textit{C4-NOT} is transpiled into a 34-gate circuit: 14 \textit{CNOT} gates, 18 $RZ$ and two $SX$. A single \textit{C4-NOT} gate thus needs a large number of gates that form a fairly large circuit by themselves, and the implementation from  \cite{GithubQW} contains two Toffolis and two \textit{C4-NOT} gates for a single step of the quantum walk. This explains why the quantum walk on \texttt{ibm\_quito} appears as just noise. The number of gates being implemented, even for a single step, has enough noise associated with each gate to degrade the results to the point where it is unrecognisable. In comparison, the 4-node circuit does not contain any Toffoli or \textit{C4-NOT} gates, so the only gates to transpile are $H$ and $X$, which explains why the results achieve a higher fidelity. This circuit, when transpiled is much closer to figure \ref{fig:4node_DTQW_circuit}, and contains a modest number of gates, so noise degradation does not ruin the result. It is worth mentioning that the transpiling process does not solely consist of transforming the circuit into gates, but also decides which qubits to use in the device, determined by the hardware's topology, among other processes \cite{transpile}. Factors such as the aforementioned topology constraints can have a negative impact on the fidelity of a computation by, for example, increasing the total number of gates in a circuit, leading to more noise.

\section{Custom Noise Models}
\label{sec:custom_noise_models}

A feature of the qiskit software package is the ability to create and use noise models. IBM offers a noise model of each available backend that, in principle, models the noise levels of that device, on that day, as accurately as qiskit will allow. However, it is also possible for the user to create their own custom noise models by varying, and including/not including, parameters associated with noise in superconducting quantum computers. Noise models are created using the \texttt{qasm\_simulator} as a base template, noise parameters are then introduced on top of this template. The main parameter in the custom noise models comes from the depolarising error channel, and is modelled by \cite{qiskit_depolarising}
\begin{equation}
E(\rho) = (1 - \lambda) \rho + \lambda \textrm{Tr} [\rho] \frac{\mathds{1}}{2^{n}}
\end{equation}
with $0 \leq \lambda \leq 4^{n}/(4^{n}-1)$, where $\lambda$ is the depolarising error parameter, $n$ is the number of qubits, $\mathds{1}$ is the identity matrix (normalized by dividing by $2^n$) and $\rho$ is the density matrix of the state.

For depolarising noise, the quantum correlations are removed over time, and this will turn a quantum walk into a classical random walk.  The limiting distribution for a classical random walk on any regular graph (such as the cycle) is a uniform distribution, independent of the initial state.  
However, we are simulating noise on the quantum computer, which is not the same as a model of a noisy quantum walk.  Nonetheless, removing quantum correlations will still have the effect over long times of making the output of the computation a random distribution over the possible outcomes, independent of the initial state.

\begin{figure}[!htb]
\centering
\includegraphics[scale=0.5]{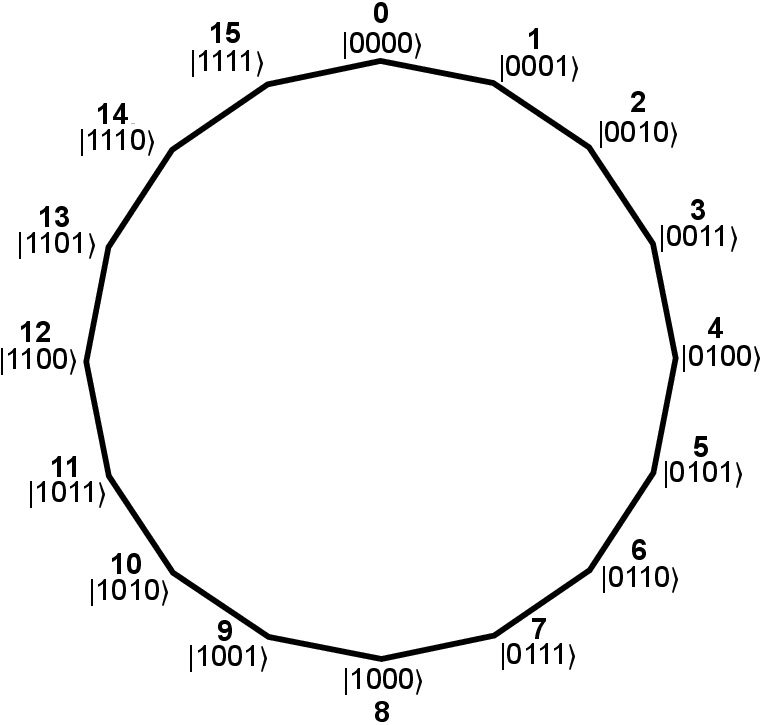}
\caption{Diagram of the 16-node cycle graph used in the DTQW. Nodes represent the possible positions of the walker and the binary string corresponds to the node label in decimal (bold) }
\label{fig:16_node_graph}
\end{figure}

\begin{figure}[!htb]
\centering
\includegraphics[width = 7.6cm, height = 2.8cm]{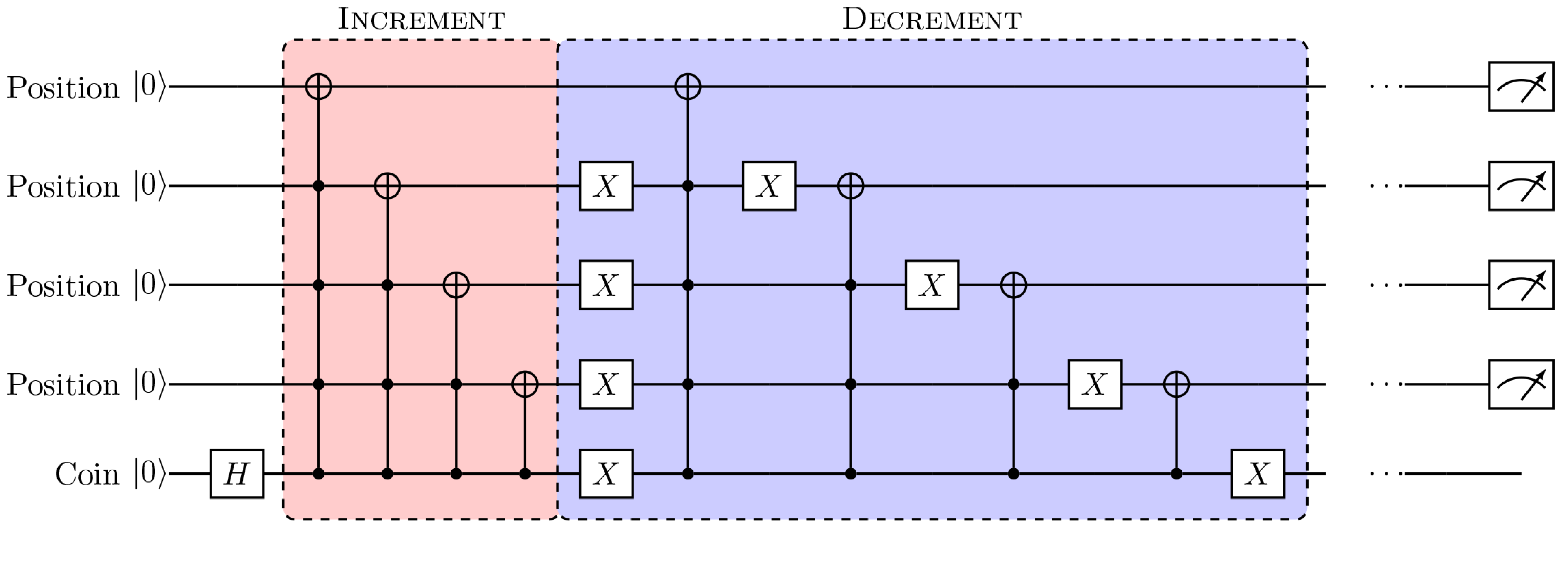}
\caption{Quantum circuit for the 16-node DTQW using 5 qubits. Figure depicts one step of the walk. Bottom most qubit (as pictured) acts as the coin and the rest of the qubits are used to represent the position of the walker.}
\label{fig:16_node_circuit}
\end{figure}

We can use the customized noise model to estimate how much noise levels in the IBM quantum device \texttt{ibmq\_santiago} would need to be reduced in order to achieve a consistent, high-fidelity result for a discrete-time quantum walk.
To investigate this, consider the DTQW on a cycle graph, using the same paradigm as the walks in the previous section. However, now the quantum register is extended to a total of five qubits, this being the maximum number available to us. Four qubits are used for the binary encoding of the position of the walker and the fifth qubit stores the coin state. Accordingly, the number of sites on the graph, for this quantum walk, can now increase to a total of 16, as shown in figure \ref{fig:16_node_graph}. The corresponding quantum circuit for this walk is shown in figure \ref{fig:16_node_circuit}.

The parameters that make-up a noise model are: single-qubit gate errors, two-qubit gate errors, three-qubit gate errors, and multiple (i.e., four and five for this algorithm) qubit gate errors. The single-qubit gate errors correspond to the single qubit unitary operators in the circuit. In the case of the 16-node DTQW that we use, this includes $H$ and $X$ gates. The two, three and multiple qubit errors correspond to \textit{CNOT}, Toffoli and \textit{C4-X}/\textit{C5-X} gates respectively. In practice, these errors represent depolarising error rates associated with each of the gates, and are numbers specified by the user.

In order to create a custom noise model that is able to accurately model a real device, we begin by an initial guess of the error rates associated with each of the aforementioned gate types, single-qubit, two-qubit etc. We then examine the outputted results of the 16-node, 16-step cycle DTQW, run on this custom noise model, and compare it to the results outputted by the real device. By examining how close the Hellinger fidelity of the two distributions (from the custom noise model and real device) are, the values for the gate errors can be refined. By refining the gate errors until a Hellinger fidelity of 1.0, or as close to that value as possible, is achieved, a custom noise model of a real device is able to be constructed. The final values for the gate errors, which are the probability of error per gate, are shown in table \ref{table:errors_table} below.

\begin{table}[!htb]
\begin{tabular}{ | P{3.5cm} | P{1.5cm}| P{1cm} | } 
  \hline
  \textit{Type of gate} & \textit{Error} \\ 
  \hline
  single-qubit & 0.005 \\ 
  \hline
  two-qubit & 0.02 \\ 
  \hline
  three-qubit & 0.04 \\ 
  \hline
  multiple-qubit & 0.6 \\
  \hline
\end{tabular}
\caption{\centering Error rate per gate used in the custom noise model to emulate \texttt{ibmq\_santiago}. Multiple-qubit refers to four or more qubit gates.}
\label{table:errors_table}
\end{table}
The values shown in table \ref{table:errors_table} are referred to in this text as \emph{full noise strength}, as they refer to the maximum amount of noise modelled i.e., the amount of noise that most closely resembles the real device.

As mentioned previously, IBM also provide a noise model for each of their backends. This noise model has the error rates pre-determined by automated analysis of each of the devices every so often e.g., every 24 hours. Due to the susceptibility of transmon qubits to environmental noise, these error rates vary over time. Hence, this method of modelling noise is a means of keeping track of that variation, and provides the most accurate error values. The other advantage is that the user does not need to define each individual error rate to match the current noise levels, this is done automatically. In addition to depolarising errors, the IBM noise model also contains: readout errors, errors associated with the measurement of qubits, and a thermal relaxation error that essentially consists of the $T_{1}$ and $T_{2}$ relaxation times \cite{IBM_NM_errors}. It is important to note that the depolarising and thermal relaxation errors only apply to single and two-qubit gates, so the noise model needs to be applied to circuits with three-qubit and higher gates transpiled into one and two qubit gates. In this paper, the noise model provided by IBM is referred to as the \emph{IBM noise model}.

In order to investigate how noise affects the performance of this generation of IBM quantum computers, the DTQW for 16 nodes is executed, first on the base \texttt{qasm\_simulator} with no noise, and second using the custom noise model, for 16 steps of the walk in single step increments. The noise level is then reduced in gradual decrements, until there are no depolarising errors in the custom noise model. The same DTQW is then executed on the IBM noise model of \texttt{ibmq\_santiago} and the real \texttt{ibmq\_santiago}. The fidelity for each step of the quantum walk is recorded up to a maximum of 16 steps. The fidelity for each backend: custom noise model, IBM noise model and \texttt{ibmq\_santiago}, is found by comparing the probability distributions of each step of the walk with the corresponding step from the \texttt{qasm\_simulator}, which can be thought of as the ideal or expected result, as it contains no noise.
\begin{figure}[!tbh]
\centering
\includegraphics[width=\columnwidth]{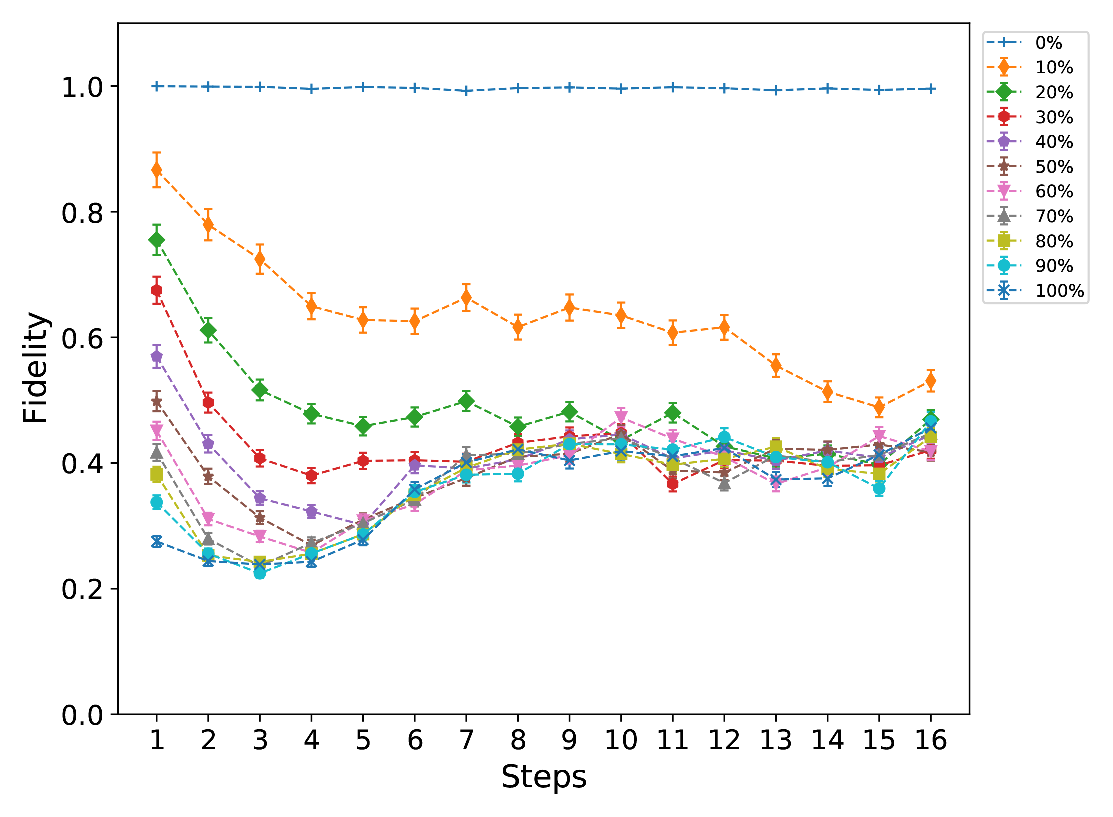}
\caption{Fidelity at each step of a 16-node, 16-step DTQW, for varying noise strengths of the custom noise model. Error bars represent standard error. }
\label{fig:CNM_strength_graph}
\end{figure}

Figure \ref{fig:CNM_strength_graph} shows the fidelity for each step of the quantum walk and for decreasing noise levels. In figure \ref{fig:CNM_strength_graph}, 100\% refers to 100\% of the full noise strength, i.e., the values in table \ref{table:errors_table}; 90\% refers to 90\% of each error in table \ref{table:errors_table}, etc. At 0\% of full noise strength the errors all take a value of 0. However, it is important to note that this is not exactly the same (although extremely close) as the base \texttt{qasm\_simulator}. The distinction is that at 0\% noise strength the noise model still takes into account the small errors introduced by the transpilation process (see section \ref{sec:transpile} and \cite{transpile} for explanation).

\begin{figure}[!tbh]
\centering
\includegraphics[width=\columnwidth]{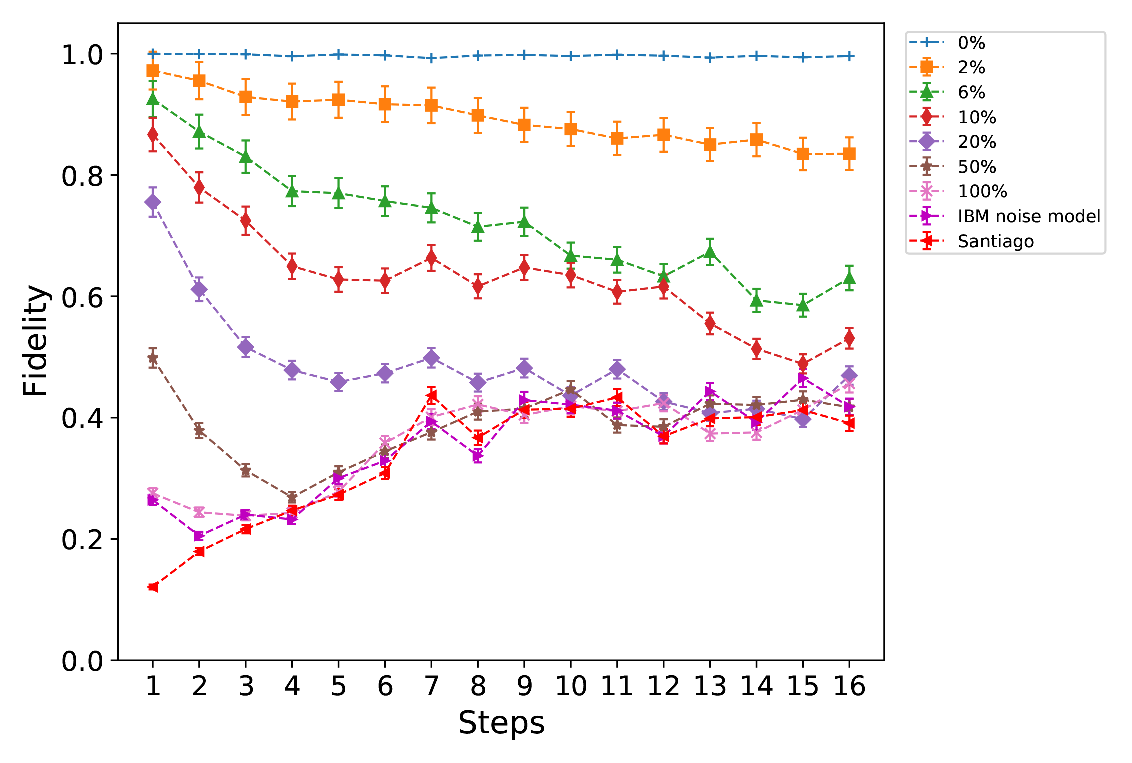}
\caption{Fidelity at each step of a 16-node, 16-step DTQW, for varying noise strengths of the custom noise model as well as the IBM noise model and \texttt{ibm\_santiago}. Error bars represent standard error.}
\label{fig:Noise_strength_graph}
\end{figure}

As can be seen from figure \ref{fig:CNM_strength_graph}, for all noise strengths except 0\%, the performance is poor. Noise strengths 60-100\% all begin with a fidelity below 0.5, in other words they cannot even produce an accurate result for one step of the quantum walk (recall that one step consists of the sequence of single and multi-qubit gates shown in Fig.~\ref{fig:16_node_circuit}). Similarly, noise strengths 30-50\% show some improvement but still not enough to be considered accurate. Significant improvements in the fidelity only begin to be seen when at 10\% and 20\% of full noise strength. However, the fidelity soon declines in the case of 20\% noise strength, and after just 2 steps the results of the walk are indiscernible from random outputs. In the case of 10\% of full noise strength, the performance is markedly better, with the fidelity being consistently higher than all other noise strengths for every step of the walk. Here, approximately 13 steps of the walk are achieved at a reasonable standard with a final fidelity around 0.6, which is roughly the limit before the results of the walk start to become indistinguishable from random outputs i.e., noise. The 0\% noise strength achieves a constant fidelity of almost exactly 1.0, with minor fluctuations occurring due to the transpilation errors. This confirms that 0\% noise strength and the \texttt{qasm\_simulator} are very similar. In all other cases for the custom noise model, it can be seen that the fidelity tends to an equilibrium value of approximately 0.5 as the number of steps tends to 16, showing the effect of noise overcoming the results of the walk.
Figure \ref{fig:Noise_strength_graph} shows the fidelity for the same DTQW, for selected noise strengths, and includes the noise model of \texttt{ibmq\_santiago} as well as \texttt{ibmq\_santiago} itself.

From figure \ref{fig:Noise_strength_graph} a direct comparison between the noise models and \texttt{ibmq\_santiago} can be made. Overall, the worst performing is the real device, starting with a low fidelity that gradually increases but never exceeds 0.5. This indicates a slight anti-correlation with the expected results, particularly for the first few steps. However, beyond roughly 6 steps these tend to random outputs (noise). The performance of the IBM noise model at 100\% strength is very similar to \texttt{ibmq\_santiago}. For the IBM noise model, this would be expected but the custom noise model also follows the real device very closely, overlapping at 4, 9 and 10 steps. Although overall the performance of the IBM noise model is slightly closer to the real device, it is interesting to see this near correspondence. This indicates that depolarising errors are the dominant contribution to noise, since it is possible to so closely model (as shown in figures \ref{fig:CNM_strength_graph} and \ref{fig:Noise_strength_graph}) the real device without the addition of other errors. The addition of 2\% and 6\% noise strength also shows the non-linear improvement in performance once the level of noise is below 10\% of the initial parameters. For 6\% noise strength, the entire walk of 16 steps is achievable with a reasonable fidelity, only dropping slightly towards 16 steps, even at this point the results of the walk are distinguishable from any random outputs (noise). The 2\% noise strength performs excellently, and by far has the highest overall fidelity (aside from 0\% of course), never dropping below 0.8 fidelity. All steps of the walk are considered accurate and extrapolation of figure \ref{fig:Noise_strength_graph} would seem to suggest that many more additional (probably around 8) steps could be achieved with high or good fidelity results for 2\% noise strength. We therefore conclude that in order for a DTQW on a 16-node cycle graph to be executed with consistent, high fidelity results, for each step, the noise levels within \texttt{ibmq\_santiago} would need to be reduced by approximately 94\%.

The explanation for the poor fidelity for the 8-node cycle presented in in section \ref{sec:execution} is now clear.  The poor fidelity of \texttt{ibmq\_santiago}, the IBM noise model and most of the custom noise models, is due to the transpilation process expanding the number of native gates, and with it, the total noise.

\section{Conclusion}
\label{sec:conclusion}

We have carried out two different size cycle discrete-time quantum walks on an IBM quantum computer, \texttt{ibmq\_quito}, showing that the device is unable to produce high fidelity results for an 8-node, 8-step walk. However, it is able to produce reasonably high fidelity results for a 4-node, 4-step walk. We have established the reason for this discrepancy in results between the two walks is due to the transpilation process. Specifically, the 4-node walk uses fewer qubits and fewer gates, as well as using only \textit{CNOT} gates, while the 8-node walk requires \textit{C3-X} (Toffoli) and \textit{C4-X} gates, that are transpiled into many more native gates. Therefore, less noise occurs in the execution of the 4-node walk, and it is able to achieve a much higher fidelity. Section \ref{sec:custom_noise_models} then established, using custom noise models, a method of approximating  by how much the noise in \texttt{ibmq\_santiago} would need to be reduced in order to execute a 16-node, 16-step cycle discrete-time quantum walk. Inspection of figures \ref{fig:CNM_strength_graph} and \ref{fig:Noise_strength_graph}, revealed that a decrease in noise levels of approximately 94\% would be sufficient to achieve this task.

Although our conclusions show that this generation of IBM quantum computers have a long way to go in terms of reduction of noise levels, for even a modest-sized DTQW, it is encouraging that a smaller DTQW is currently viable.

Quantum walks are by no means the only algorithm that can be used to benchmark performance of quantum computers. For example, randomised benchmarking \cite{Randomized_benchmarking1, Randomized_benchmarking2} has been used to characterise noise in IBM quantum computers. Furthermore, combinatorial optimisation problems have also been used to measure performance \cite{optimisation_benchmark}. However, fundamentally, all quantum algorithms implemented on IBM superconducting quantum computers use quantum gates. Specifically, we tested the gates in section \ref{sec:transpile}. Therefore, the conclusions regarding transpilation and noise associated with each gate, highlighting in particular Toffoli, \textit{C4-X} and \textit{C5-X} gates, hold true for any algorithm that uses those gates, implemented on these processors.

\subsection*{Acknowledgements}

VK and NC were partially funded by UKRI EPSRC Grant No. EP/T026715/1 and EP/T026715/2, and by the UK Quantum Technology Hub in Computing and Simulation (grant EP/T001062/1).

We acknowledge the use of IBM Quantum services for this work. The views expressed are those of the authors, and do not reflect the official policy or position of IBM or the IBM Quantum team.

\subsection*{Declarations}
\textbf{Competing Interests} The authors have no relevant financial or non-financial interests to disclose.

\noindent\textbf{Author contributions} VK and NC proposed initial study. VK proposed use of fidelity. VW designed 4-node DTQW circuit and custom noise models. VW performed data collection. All authors performed analysis. First draft of the manuscript was written by VW, VK and NC contributed to subsequent amendments. All authors read and approved the final manuscript.  

\noindent\textbf{Data Availability Statement} Datasets and code used in this manuscript available at the Durham University Collection: \url{http://doi.org/10.15128/r15x21tf47n}


%

\end{document}